# ChemCloud: Chemical e-Science Information Cloud

Alexandru Todor, Adrian Paschke

Freie Universität Berlin

Stephan Heineke

FIZ Chemie Berlin

**Abstract:** Our Chemical e-Science Information Cloud (ChemCloud) – a Semantic Web based eScience infrastructure – integrates and automates a multitude of databases, tools and services in the domain of chemistry, pharmacy and bio-chemistry available at the Fachinformationszentrum Chemie (FIZ Chemie), at the Freie Universitaet Berlin (FUB), and on the public Web. Based on the approach of the W3C Linked Open Data initiative and the W3C Semantic Web technologies for ontologies and rules it semantically links and integrates knowledge from our W3C HCLS knowledge base hosted at the FUB, our multi-domain knowledge base DBpedia (Deutschland) implemented at FUB, which is extracted from Wikipedia (De) providing a public semantic resource for chemistry, and our well-established databases at FIZ Chemie such as ChemInform for organic reaction data, InfoTherm the leading source for thermophysical data, Chemisches Zentralblatt, the complete chemistry knowledge from 1830 to 1969, and ChemgaPedia the largest and most frequented e-Learning platform for Chemistry and related sciences in German language.

## 1  Introduction

Modern scientific research requires vast amounts of data in order to achieve significant results. Most of this data comes from a variety of heterogeneous knowledge bases, and requires a great amount of manual work in order to be linked together in a scientifically meaningful way. This problem is especially relevant in the Chemistry and Life Sciences domains, where most knowledge is stored in thousands of databases and many millions of scientific publications [1]. Adding to this problem are vast amounts of experimental data produced every day in laboratories around the world. This regular flood of data leads in most cases to an inefficient use of information where many valuable data sources are ignored, slowing or hampering the study or research activity.
ChemCloud tries to overcome this problem by using a Semantic Data Integration approach in order to bridge the gap between multiple databases that span across different knowledge domains [2]. In contrast to traditional data integration approaches, Semantic Data Integration is more flexible and also offers the possibility of capturing the richness of relationships between concepts coming from multiple sources. The use of semantics allows us to capture its meaning of a concept rather than its pure syntactical significance, which proves very especially important in scientific texts where the same word can have different meaning based on context or domain [3].

## 2 Architecture

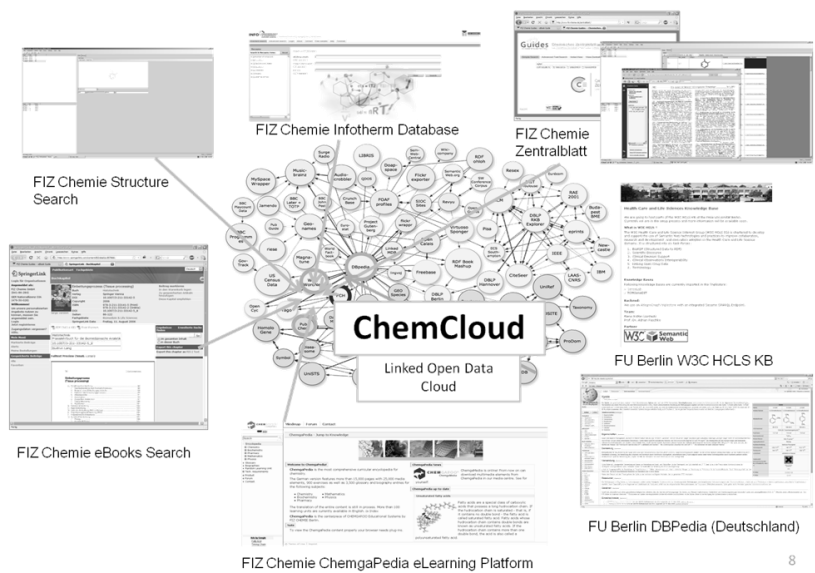

ChemCloud is composed of multiple independent data projects, the current main focus lies on two databases offered by Fiz-Chemie: InfoTherm, a thermphisical and thermochemical properties database containing data on more than 23 thousand compounds and ChemgaPedia, the largest eLearning platform for chemistry in Germany. The main objective is the rdfization of these databases and the semantic integration with the W3C HCLS KB and DBpedia Germany.

In order to overcome the hurdles posed by the different data models and data formats from the various databases in ChemCloud we rely RDF as a common language to enable widespread aggregation and reuse of the data via publication as Linked Data on the web [4]. Furthermore machine processable ontologies are used for providing the semantic mappings between related terms and to process, integrate, and reason across the data retrieved from separate databases [5].

## 3 Ontological Model

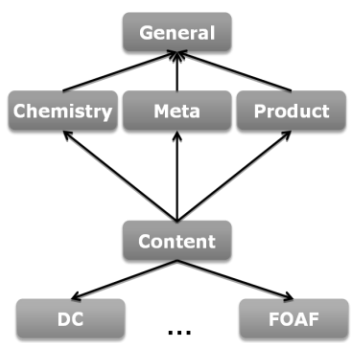

The general ontological model of ChemCloud is composed of a set of modular ontologies:General ontologies cover more common concepts from the application domain such as content, product and term, while specialized domain ontologies describe concepts such as thermophysical and thermochemical properties, reactions and compound classes.

The product ontologies focus solely on describing and classifying the products and services offered by FIZ-Chemie, and the meta domain ontologies describe things such as provenance of data, eBooks, eLearning concepts and contain mapping ontologies.

Additionally we use a series of standard ontologies in our ChemCloud ontology model such as FOAF(Persons) and Dublin Core(content).

## 4 Data Integration

Due to the strong reliance on XML technologies in the existing databases we transitioned from XML to Semantic Web standards, keeping the underlying semantics in the XML schema and enriching the data model with semantics from other knowledge bases.

Although XML is an efficient format for storing structured data, it is not suitable for complex data integration purposes. XML has very limited abilities for achieving the interoperability required by complex applications that span multiple science domains. XML is document centric and more of a syntactic than a semantic nature., it uses tags in order to structure the data inside the documents, but those tags do little to describe the data they contain.

The XML based chemical databases are focused only on a specific XML schema (i.e. ThermoML in the case of InfoTherm), however, a large amount of data is described by other XML based chemical markup languages like CML(chemical markup language), MatML (materials markup language), AniML (Analytical Markup Language) and many other databases that describe all sorts of chemical, biochemical and biological data. The proliferation and redundancy of the XML formats in the chemistry and life sciences domains poses major problems towards data integration. The formats described above, although originally developed with interoperability in mind, have ended up describing the same or related data, and XML offers no efficient mechanism of gluing different formats together in order to achieve the requirements of a complex data integration project.

### 4.1 Conversion

In order to achieve semantic integration, we need to transform the XML data into Semantic Web formats. The first step in this process is to create ontologies that accurately describe the data in the XML documents. This transition is a complex task since the most of the undelying XML schemas describe a large data model that is difficult to represent in an ontology. Due to this complexity we have split the conversion process into two parts, syntactic and semantic conversion [6].

Semantic conversion requires the propper ontological conceptualization of the domain knowledge. This process includes the development of the ChemCloud ontological model as well as the alignment with other compatible ontologies.

For better compatibility with established knowledge bases we need to align our ontological model with existing chemical ontologies such as: CheBI (Chemical Enteties of Biological Interest) [7], a detailed ontology that covers a large part of the chemistry domain and ChemAxiom [8], an ontological framework that among other things focuses strongly on chemical properties. The Bio2RDF [9] ontology is also important subject for alignment since this knowledge base contains vast amounts of information.

The alignment process for chemical compounds is usually straight forward since the properties that describe them (names, formulas, identifiers etc.) are quite similar in all ontologies; however aligning thermophysical and thermochemical properties classes and measurement methods proves more complicated.

Syntactic conversion in contrast is a more streight-forward approach that can be done trough semi-automatic means. This process presumes the semi-automatic creation of an ontology derived from the XML schema, in order to faccilitate the conversion of the XML data into RDF. In parallel to the syntactic conversion process, semantic conversion is undertaken by mapping the ontology derived from the XML schema to expert-made ontologies, that describe the domain of interest in a more detailed manner.

### 4.2 Data Source Integration

The current Linked Open Data web offers a wide variety of chemical and biochemical databases that contain vast amounts of scientific data. RDF allows us to enhance ChemCloud with complementary information by leveraging databases from Bio2RDF such as Chebi and Pubchem. It enables us to add synonyms from different languages, other chemical identifiers, compound roles in different solutions or uses in different industries. Integrating with DBpedia (more specifically DBpedia Germany) further allows us to better classify the chemical concepts based on the SKOS categorization and to add multilingual descriptions of the different compounds.

### 4.3 New Features

Data Integration with existing semantic knowledge bases enables us to make a series of new functionality available to for existing chemical databases such as:

Semantic autocomplete: in order to avoid misspellings and ease the search for chemical compounds by name, making use of the synonyms offered by different databases from BIO2RDF. This allows us to predict what terms from the vocabulary the user is trying to search for. The proposed terms can be selected and thereby misspellings are avoided and search process is simplified by not having to type the entire keywords.

We can also find synonyms based on the relations between terms in our vocabulary or from different languages by matching the keywords expressed in other languages to the same concept.

Semantic faceted search: allows the user to form queries in a more simple and guided way by picking predefined search filters from the ontology. The user can limit the search domain by selecting new filters step by step in an iterative process until reaching the desired result. New filters are suggested at each step based on the ontological hierarchy. Classical keyword based searches with a hardcoded list of filters are usually based only on syntactical matching, they are sensible to mistyping and will only give results for exact matches. Ontology based faceted search, gives us the possibility to also return related ontological choices from the ontology, even if they do not syntactically match the typed keyword.

## 5   Conclusions and Future Work

In this poster paper we have described ChemCloud which integrates several high quality databases of FIZ - Chemie in the linked data cloud. It is an ongoing project that aims to bring together several isolated knowledge bases, create a robust model for the transition to Linked Data of existing chemical databases and enable new features and research possibilities trough data integration and semantic technologies such as: semantic search, semantic data enrichment, ontology-enhanced navigation, automatically generated eLearning trajectories and semantic knowledge discovery over multiple databases.